\language1
\count0=1
\def\nb #1{{\hbox{\bf #1}}}
\font\bsym=cmbsy10
\def\bnabla{\hbox{$\textfont2=\bsym \nabla$}}
\def\bscal#1{\hbox{$\textfont1=\boit #1$}}

\font\boit=cmmib10 scaled\magstephalf
\font\huge=cmr12
\font\biggb=cmbx12 scaled\magstep1
\font\ninep=cmr8
\font\tenp=cmr10
\font\twelvebf=cmbx12

\hsize=15truecm
\vsize=20.6truecm
\hoffset=0.6 truecm
\tolerance=10000
\hbadness=10000
\hfuzz=20pt
\baselineskip 15pt
\footline={\ifnum\pageno=1
              \relax
           \else
\ifodd\pageno\rightfootline
              \else\leftfootline
              \fi
           \fi}
\voffset=3\baselineskip
\def\leftfootline{\hfill{\bf\folio}\hfill
}
\def\rightfootline{\hfill
 \bf{\folio}\hfill}

\vbox{\vskip .5truecm}
\noindent
{\biggb Dynamics of topological magnetic solitons}\footnote*{\ninep to
appear in the proceedings of the workshop on Solitons, Kingston (1997)}

\bigskip
\bigskip

{\huge N. Papanicolaou}

{\tenp Department of Physics, University of Crete,}

{\tenp and Research Center of Crete,}

{\tenp Heraklion, Greece}

\vbox{\vskip 1truecm}

\noindent
{\bf Abstract}
\bigskip
{\tenp A direct link between the topological complexity of magnetic
media and their dynamics is established through the construction of
unambiguous conservation laws for the linear and angular momenta as
moments of a topological vorticity. As a consequence, the dynamics of 
topological magnetic solitons is shown to exhibit the characteristic
features of the Hall effect of electrodynamics or the Magnus effect of 
fluid dynamics. The main points of this program are reviewed here for both
ferromagnets and antiferromagnets, while a straightforward extension to the
study of superfluids is also discussed briefly.}

\font\stlett=cmr12
\stlett
\bigskip
\bigskip
\noindent
{\twelvebf 1.$\,\,$ Introduction}
\bigskip
Ferromagnetic (FM) bubbles are the best known examples of magnetic solitons
and exhibit some distinct dynamical features due to their nontrivial 
topological structure. The inherent link between topology and dynamics
was already apparent in the early work of Thiele [1] as well as in many
investigations that followed [2]. The essence of the early work is best
summarized by the experimentally observed skew deflection of FM bubbles
under the influence of an applied magnetic-field gradient. The 
so-called golden rule of bubble dynamics relates the deflection
angle $\delta$ to the winding number $Q$ by
$${gr^2\over 2V}\sin\delta =Q,\eqno (1)$$
where $g$ is the strength of the applied field gradient, $r$ is the bubble
radius, and $V$ its speed. This relation is remarkable in two respects.
First, it suggests that only topologically trivial $(Q=0)$ bubbles move
in the direction of the gradient $(\delta =0)$, even though such a 
behavior would naively be expected for all FM bubbles; in fact,
bubbles with a nonvanishing winding number $(Q=\pm 1, \pm 2, \ldots)$
tend to be deflected in a direction nearly perpendicular
$(\delta\sim\pm 90^{\circ})$ to the applied gradient, exactly so in the 
limit of vanishing dissipation. Second, eq. (1) implies some sort of a 
topological quantization in that it relates the integer-valued winding
number to experimentally measured quantities that can, in principle,
assume any values.

A fresh look at this problem was initiated in ref. [3] where the link
between topology and dynamics was made explicit by the construction of 
unambiguous conservation laws as moments of a suitable topological vorticity.
The important qualitative features of bubble dynamics became then apparent.
Thus, in the absence of external magnetic-field gradients or other
perturbations, a bubble with a nonvanishing winding number cannot move
freely but is always spontaneously pinned. On the other hand, in the
absence of dissipation, a bubble with $Q\not =0$ would be deflected at
a right angle $(\delta =\pm 90^{\circ})$ with respect to an applied
gradient, while its drift velocity can be calculated analytically in some
important special cases and is generally consistent with eq. (1). The 
emerging picture is completely analogous to the Hall motion of an electron
as well as to the Magnus effect of fluid dynamics. These analogies further
suggest that the deflection angle should deviate from $90^{\circ}$
in the presence of dissipation. However an exact calculation of the 
deflection angle, i.e., a rigorous derivation of the golden rule, is no
longer possible on the basis of conservation laws alone.
Nonetheless the theoretical picture can be completed with more or less
straightforward numerical simulations.

In this short review the emphasis is placed on some general dynamical 
features that enable one to detect in a systematic manner the existence
(absence) of Hall or Magnus behavior in any field theory that bears
topological solitons. The general framework is then briefly illustrated
for ferromagnets (FM), antiferromagnets (AFM), and superfluids.
In the process, we provide a complete list of references to the recent
work where the issues involved are discussed in greater detail.

\bigskip
\bigskip
\noindent
{\twelvebf 2.$\,\,$ Vorticity and conservation laws}
\bigskip
In this section we study a general field theory governed by the 
Hamilton equations
$$\dot{\Psi}_a={\delta W\over \delta\Pi_a},\qquad
\dot{\Pi}_a=-{\delta W\over \delta\Psi_a},\eqno (2)$$
where $W$ is the Hamiltonian and $(\Psi_a, \Pi_a)$, with
$a=1, 2, \ldots N$, is a set of $N$ canonically conjugate pairs of fields
satisfying the standard Poisson bracket relations. We consider first a
strictly two-dimensional (2D) theory defined in a plane with 
coordinates $\nb x=(x_1, x_2)$. One may then construct the scalar density
$$\gamma =\varepsilon_{\mu\nu}\partial_{\mu}\Pi_a\partial_{\nu}\Psi_a,
\eqno (3)$$
where each of the Greek indices $\mu$ and $\nu$ is summed over two
distinct values corresponding to the two spatial coordinates $x_1$ and $x_2$,
$\varepsilon_{\mu\nu}$ is the 2D antisymmetric tensor, and the index
$a$ is summed over all $N$ values counting the number of canonical pairs.
The density $\gamma$ will be referred to as vorticity because it shares
several formal properties with ordinary vorticity in fluid dynamics.
The time derivative of the vorticity is then calculated from the
Hamilton equations (2) to yield
$$\dot{\gamma}=-\varepsilon_{\mu\nu}\partial_{\mu}\tau_{\nu},\eqno (4)$$
where the vector density
$$\tau_{\nu}={\delta W\over \delta\Psi_a}\partial_{\nu}\Psi_a+
{\delta W\over \delta\Pi_a}\partial_{\nu}\Pi_a,\eqno (5)$$
is formally analogous to the ``force density'' employed by Thiele [1]
in the problem of FM bubbles. We proceed one step farther noting that
$\tau_{\nu}$ may be written as a total dirergence,
$$\tau_{\nu}=\partial_{\lambda}\sigma_{\nu\lambda},\eqno (6)$$
where $\sigma_{\nu\lambda}$ is the stress tensor
$$\sigma_{\nu\lambda}=w\delta_{\nu\lambda}-
{\partial w\over\partial (\partial_{\lambda}\Psi_a)}\partial_{\nu}\Psi_a-
{\partial w\over\partial (\partial_{\lambda}\Pi_a)}\partial_{\nu}\Pi_a
\eqno (7)$$
defined in terms of the energy density $w$ identified from 
$W=\int wdx_1dx_2$. Equation (4) then becomes
$$\dot{\gamma}=-\varepsilon_{\mu\nu}\partial_{\mu}\partial_{\lambda}
\sigma_{\nu\lambda}\eqno (8)$$
and proves to be fundamental for our purposes [3].

It should be noted that the preceding discussion makes no distinction
between ordinary field theories and those endowed with nontrivial topological
structure or related properties. However a clear distinction emerges when
we consider the total vorticity
$$\Gamma =\int \gamma dx_1dx_2=\varepsilon_{\mu\nu}\int\partial_{\mu}
\Pi_a\partial_{\nu}\Psi_a dx_1dx_2,\eqno (9)$$
which is conserved by virtue of eq. (8) for any field configuration
with reasonable behavior at infinity. One may also write
$$\Gamma =\varepsilon_{\mu\nu}\int [\partial_{\mu}
(\Pi_a\partial_{\nu}\Psi_a)-\Pi_a\partial_{\mu}\partial_{\nu}\Psi_a]
dx_1dx_2\eqno (10)$$
to indicate that a vanishing value of the total vorticity is the rule
rather than the exception. Indeed, under normal circumstances, the
first term in eq.$\!$ (10) is shown to vanish by transforming it into a
surface integral at infinity and the second term also vanishes
because $\varepsilon_{\mu\nu}\partial_{\mu}\partial_{\nu}\Psi_a=0$
for any differentiable function $\Psi_a$. Yet the above conditions may
not be met in a field theory with nontrivial topology, thus leading to
ambiguities in the canonical definitions of linear momentum 
$\nb p=(p_1, p_2)$ and angular momentum $\ell$ given by
$$p_{\mu}=-\int \Pi_a\partial_{\mu}\Psi_a dx_1dx_2,\qquad
\ell =-\int \Pi_a\varepsilon_{\mu\nu}x_{\mu}\partial_{\nu}\Psi_adx_1dx_2.
\eqno (11)$$
In general, the above canonical conservation laws are rendered
ambiguous when the total vorticity $\Gamma$ is different from zero.

Nevertheless unambiguous conservation laws can be constructed by
returning to the fundamental relation (8) where the appearance of a
double spatial derivative in the right-hand side implies that some
of the low moments of the local vorticity $\gamma$ are conserved.
Indeed the linear momentum is given by
$$p_{\mu}=-\varepsilon_{\mu\nu}I_{\nu},\qquad
I_{\nu}\equiv\int x_{\nu}\gamma dx_1dx_2,\eqno (12)$$
and the angular momentum by
$$\ell ={1\over 2}\int\rho^2\gamma dx_1dx_2,\eqno (13)$$
where $\rho^2=x_1^2+x_2^2$. The preceding identifications are made 
plausible by inserting the general expression for the vorticity
of eq. (3) in eqs. (12) and (13) and by freely performing partial
integrations to recover the canonical forms of linear and angular
momenta quoted in eq. (11). However partial integrations should be
performed with great care and are often unjustified when $\Gamma\not =0$.

The effect of a nonvanishing total vorticity becomes apparent by 
considering the transformation of the moments $I_{\nu}$ of eq. (12)
under a translation of coordinates $\nb x\to\nb x+\nb c$ where
$\nb c=(c_1, c_2)$ is a constant vector:
$$I_{\nu}\to I_{\nu}+\Gamma c_{\nu},\eqno (14)$$
which implies a nontrivial transformation of the linear momentum
(12) when $\Gamma\not =0$. This is surely an unusual property and
indicates that the moments $I_{\nu}$ provide a measure of position
rather than momentum. Such a fact is made explicit by considering the
guiding-center vector $\nb R=(R_1, R_2)$ with coordinates
$$R_{\nu}={I_{\nu}\over\Gamma}={1\over\Gamma}\int x_{\nu}\gamma
dx_1dx_2,\eqno (15)$$
which transforms as $\nb R\to\nb R+\nb c$ under a constant translation
and is thus a measure of position of a field configuaration with
$\Gamma\not =0$. Nevertheless, the vector $\nb R$ is conserved.

A related fact is that the familiar Poisson bracket algebra is 
significantly affected when $\Gamma\not =0$. Using the canonical
Poisson brackets
$$\{\Pi_a(\nb x), \Psi_b(\nb x')\} =\delta_{ab}
\delta (\nb x-\nb x')\eqno (16)$$
and the general expression of the local vorticity (3) in the definition
of the linear momentum (12), it is not difficult to establish the relations
$$\{ p_1, p_2\} =\Gamma,\qquad \{ R_1, R_2\} =1/\Gamma ,\eqno (17)$$
which are strongly reminiscent of the situation in the case of electron 
motion in a uniform magnetic field, the role of the latter being 
played here by the total vorticity $\Gamma$.

Similarly, the angular momentum (13) actually provides a measure of the
soliton size, a fact made explicit by considering the mean squared
radius defined from
$$r^2={1\over\Gamma}\int (\nb x-\nb R)^2\gamma dx_1dx_2=
{2\ell\over\Gamma}-\nb R^2,\eqno (18)$$
which is also conserved. Needless to say, the conservation laws (12) and 
(13) resume their ordinary physical significance at vanishing total
vorticity $(\Gamma =0)$.

The observed transmutation in the physical significance of the 
conservation laws of linear and angular momenta implies a radical change
in the dynamical behavior. For instance, a single soliton with
$\Gamma\not =0$ cannot be found in free translational motion
$(\dot{\nb R}=0)$. It is always spontaneously pinned or frozen within
the medium, whereas translation invariance is preserved by the fact that
spontaneous pinning can occur anywhere in the $(x_1, x_2)$ plane.
Soliton motion is possible in the presence of external field
gradients or other solitons, but the dynamical pattern is also expected
to be unusual in that motion tends to take place in a direction
perpendicular to the applied force. In other words, solitons with a 
nonvanishing total vorticity are expected to behave as electric charges
in a uniform magnetic field or as ordinary vortices in a fluid. But one
should keep in mind that a topological soliton does not necessarily carry
a nonvanishing total vorticity. This and related issues can be settled
only within a definite dynamical model, as discussed further in 
subsequent sections.

This section is completed with a brief discussion of a 3D generalization.
Thus eq. (8) becomes 
$$\dot{\gamma}_i=-\varepsilon_{ijk}\partial_j\partial_{\ell}
\sigma_{k\ell},\eqno (19)$$
where Latin indices assume three distinct values and
$\varepsilon_{ijk}$ is the 3D antisymmetric tensor. The stress tensor
$\sigma_{k\ell}$ is obtained by an obvious 3D extension of eq. (7)
and the vorticity is now a vector density $\bscal{\gamma}=
(\gamma_1, \gamma_2, \gamma_3)$ given by
$$\gamma_i=\varepsilon_{ijk}\partial_j\Pi_a\partial_k\Psi_a.\eqno (20)$$
Accordingly the conserved linear and angular momenta read
$$\nb p=-{1\over 2}\int (\nb r\times \bscal{\gamma})dV,\qquad
\bscal{\ell}=-{1\over 3}\int [\nb r\times (\nb r\times\bscal{\gamma})]dV,
\eqno (21)$$
where $\nb r=(x_1, x_2, x_3)$ and $dV=dx_1dx_2dx_3$. Again, if partial
integrations are freely performed, eqs. (21) reduce to the standard
(canonical) conservation laws at $D=3$. However such integrations
may not be justified for 3D field configurations with a nontrivial 
topology; e.g., configurations with a nonvanishing Hopf index $[3, 4]$.

Finally we mention that the conservation laws (12) and (21) are formally
identical to those derived in fluid dynamics, at least for incompressible
fluids; see sect. 7 of ref. [5].

\bigskip
\bigskip
\noindent
{\twelvebf 3.$\,\,$ Ferromagnets}
\bigskip
A ferromagnetic medium is described in terms of the density of magnetic
moment or magnetization $\nb m=(m_1, m_2, m_3)$ which is generally
some function of position and time but has nearly constant magnitude for
temperatures sufficiently below the Curie point. The dynamics is governed
by the Landau-Lifshitz equation
$$\dot{\nb m}=\nb m\times\nb f;\qquad
\nb f=-{\delta W\over \delta\nb m},\qquad \nb m^2=1,\eqno (22)$$
where $W=W(\nb m)$ is a suitable energy functional and the constant
magnitude of the magnetization is normalized to unity.

We first discuss some general features of the Landau-Lifshitz equation
that do not depend on the details of the energy functional. For example,
eq. (22) may be brought to the standard Hamiltonian form of eq. (2)
by resolving the constraint $\nb m^2=1$ through, say, the spherical 
parametrization $m_1=\sin\Theta\cos\Phi$, $m_2=\sin\Theta\sin\Phi$,
$m_3=\cos\Theta$. It is then not difficult to show that eq. (22) may be
written in the form (2) with a single pair of canonical variables
$\Psi =\Phi$ and $\Pi =\cos\Theta$. For a strictly 2D theory the vorticity
(3) reads
$$\gamma =\varepsilon_{\mu\nu}\partial_{\mu}\Pi\partial_{\nu}\Psi =
-\varepsilon_{\mu\nu}\sin\Theta\partial_{\mu}\Theta\partial_{\nu}\Phi =-
{1\over 2}\varepsilon_{\mu\nu}(\partial_{\mu}\nb m\times\partial_{\nu}\nb m)
\cdot \nb m,\eqno (23)$$
where we recognize, in the third step, the familiar Pontryagin density.
We further restrict our attention to field configurations 
$\nb m=\nb m(\nb x, t)$ that approach the ferromagnetic ground state
$(0, 0, 1)$ at spatial infinity. Under such conditions the total
vorticity of eq. (9) yields
$$\Gamma =4\pi Q,\eqno (24)$$
where $Q=0, \pm 1, \pm 2, \ldots$ is the integer-valued Pontryagin
index or winding number. The FM bubbles alluded to in the Introduction
are static solutions of the Landau-Lifshitz equation with a definite
winding number. An abundance of such bubbles have been observed in practice
[2] with values of the winding number ranging up to $|Q|\sim 100$.

Therefore the current theory is a good example of application of the
general theory developed in the preceding section. An immediate
conclusion is that FM bubbles with $Q\not =0$ cannot move freely but are
always frozen within the ferromagnetic medium, in analogy with the 2D
cyclotron motion of electrons in a uniform magnetic field. However bubble
motion can occur either in the presence of external field gradients,
which break translation invariance and hence the conservation of the 
guiding center $\nb R$, or in the presence of other bubbles. We
consider the two possibilities in turn.

Let us first assume that a static bubble with winding number $Q$ is 
subjected to an external field $\nb h=(0, 0, h)$ which is turned on 
at $t=0$. The relevant dynamical question is then to predict the response
of the bubble and could, in principle, be settled by solving the 
initial-value problem for eq. (22) extended according to 
$\nb f\to\nb f+\nb h$ to include the applied field. However much can be
said without actually solving the complete initial-value problem, taking
advantage of the special nature of the conservation laws discussed in the
preceding section. Hence, if the applied field were spatially uniform,
translation invariance would be preserved and the guiding center of the 
bubble would remain fixed, even though finer details may acquire a 
nontrivial time dependence. A more interesting situation arises when the
applied field is not uniform but is some prescribed function of position
and time; i.e., $h=h(\nb x, t)$. The linear momentum (12) is no longer
conserved but satisfies the implicit evolution equation [3]
$$\dot{p}_{\mu}=\int (\partial_{\mu}h)(m_3-1)dx_1dx_2.\eqno (25)$$
The essential point is made apparent in the case of a highly idealized
field $h=gx_1$, where $g$ is a spatially uniform gradient that may
still depend on time. Then eq. (25) applied for $\mu =1$ and 2 yields
$$\dot{p}_1=m g,\qquad \dot{p}_2=0;\qquad\qquad
m \equiv\int (m_3-1)dx_1dx_2,\eqno (26)$$
where $m$ is the total magnetization in the third direction, after
subtracting off its trivial ground-state value. Equations (26) may be 
thought of as Newton's law and would suggest that the bubble moves in
the direction opposite to the gradient, also taking into account that
$m <0$. However this apparently straightforward conclusion is in 
sharp disagreement with the experimental fact that a bubble with
$Q\not =0$ actually moves in a direction perpendicular to the gradient.

On the other hand, the analogy with the Hall effect implied by the 
structure of the conservation laws suggests that a proper interpretation
of eq. (25), when $Q\not =0$, should proceed through the guiding-center
coordinates of eq. (15) which are related to the linear momentum (12)
simply by $R_{\mu}=\varepsilon_{\mu\nu}p_{\nu}/4\pi Q$.
Equation (15) is then rewritten in terms of the drift velocity
$\nb V=\dot{\nb R}$ to yield
$$V_{\mu}=\dot{R}_{\mu}={1\over 4\pi Q}\int
(\varepsilon_{\mu\nu}\partial_{\nu}h)(m_3-1)dx_1dx_2,\eqno (27)$$
or, in the case of a uniform gradient,
$$V_1=0,\qquad V_2=-{m g\over 4\pi Q}.\eqno (28)$$
The net conclusion is that the guiding center is indeed deflected at a 
right angle to the gradient irrespective of the time evolution of the
finer details of the bubble. Furthermore the expression for the drift
velocity (28) is fairly explicit and depends on the structural details
of the bubble only through its total magnetization $m$.

In order to obtain a completely explicit expression for the drift
velocity we must consider a specific form for the energy functional.
For purposes of illustration we consider the 2D isotropic Heisenberg
model defined by the Hamiltonian.
$$W=\int wdx_1dx_2,\qquad w={1\over 2}(\partial_{\mu}\nb m\cdot
\partial_{\mu}\nb m),\eqno (29)$$
which leads to an effective field $\nb f=\Delta\nb m$ in eq. (22),
where $\Delta$ is the 2D Laplacian. For completeness, we also quote
an explicit expression for the stress tensor (7), namely
$$\sigma_{\nu\lambda}=w\delta_{\nu\lambda}-(\partial_{\nu}\nb m\cdot
\partial_{\lambda}\nb m).\eqno (30)$$
Static solutions of this model coincide with the Belavin-Polyakov
(BP) instantons of the 2D Euclidean nonlinear $\sigma$ model [6].
Here we consider the special class of BP instantons given by
$$\Omega\equiv {m_1+im_2\over 1+m_3}=
\left({\overline{a}\over\overline{z}}\right)^n
\quad\hbox{or}\quad\left({a\over z}\right)^n,
\eqno (31)$$
where $z=x_1+ix_2$, $a$ is an arbitrary complex constant, and $n$ is a 
positive integer. The special solutions (31) closely resemble realistic
FM bubbles with winding numbers $Q=n$ or $-n$. We then calculate the radius
$r$ from eq. (18) and the total moment $m$ from eq. (26) to obtain
$$r^2=[(\pi/Q)\hbox{cosec}(\pi/Q)]|a|^2,\qquad m =-2\pi r^2.\eqno (32)$$
The first relation indicates that the current definition of the bubble
radius differs significantly from the naive definition $r=|a|$,
except in the limit $|Q|\to\infty$ which is sometimes referred to as the 
adiabatic limit. The second relation in (32) possesses a simple geometrical
significance, for it could also be obtained by considering a crude model
of a bubble in which the magnetization points toward the north pole,
$\nb m=(0, 0, 1)$, for $\rho >r$ and toward the south pole, 
$\nb m=(0, 0, -1)$, for $\rho <r$.

Now the total moment $m$ is independently conserved in the isotropic
Heisenberg model. We may then replace $m$ in eq. (28) by its initial
value (32) to obtain a completely explicit, as well as exact, result for
the drift velocity:
$$V_1=0,\qquad V_2={gr^2\over 2Q},\eqno (33)$$
which is consistent with the golden rule (1) applied for $V=|V_2|$
and $\delta =\pm 90^{\circ}$, as is appropriate in the absence of 
dissipation. Note, however, that eq. (33) contains the sophisticated
bubble radius of eq. (32) instead of the naive radius $r=|a|$.

The above results were confirmed in detail by a direct numerical 
simulation [7]. The effect of dissipation was also studied in models
of increasing complexity $[7, 8]$. The general conclusion was that golden
rule (1) is correct in its gross features but not in its finer details.

The dynamics of interacting FM bubbles was also studied in refs. [8].
When two bubbles are brought to a relative distance $d$, one would
naively expect that they would either converge toward each other or
move off to infinity depending on whether their interaction potential
is attractive or repulsive. In fact, two FM bubbles with like
winding numbers orbit around each other, in analogy with the 2D motion
of two interacting electrons in a uniform magnetic field or two like
vortices in a fluid. Furthermore, two FM bubbles with opposite winding
numbers undergo motion along roughly parallel trajectories, also in 
analogy with the 2D motion of an electron-positron pair in a uniform
magnetic field or the familiar Kelvin motion of a vortex-antivortex
pair in a fluid. Hence the analogy with the Hall effect of electrodynamics
or the Magnus effect of fluid dynamics is essentially complete.

This brief discussion of ferromagnetic solitons is completed with two
remarks. First, in practice, FM bubbles are not strictly 2D solitons
but occur in thin ferromagnetic films $[1, 2]$. However the current 
theoretical framework has been generalized to account for the 
quasi-2D nature of a film, with due caution on boundary effects 
including the effect of the long-range magnetostatic field [9].
Second, FM solitons of a different nature may occur in a strictly
3D ferromagnet and are characterized by the Hopf index. The potential
implications of the present framework for Hopf solitons were discussed
in ref. [4].

\bigskip
\bigskip
\noindent
{\twelvebf 4.$\,\,$ Antiferromagnets}
\bigskip
Although direct experimental evidence for the existence of topological
solitons in antiferromagnets is limited at this point, theoretical
arguments suggest that static AFM solitons should exist for essentially
the same reason as in ferromagnets. However their dynamics is significantly
different and is now governed by suitable extensions of the relativistic
nonlinear $\sigma$ model instead of the Landau-Lifshitz equation.
The relevance of the $\sigma$ model became apparent through standard
hydrodynamic approaches [10-12] but detailed applications to soliton
dynamics were carried out mostly in the Soviet literature reviewed in part
in ref. [13]. We were thus sufficiently motivated to extend the 
preceding analysis to the case of layered or 2D antiferromagnets whose
significance has increased in recent years in connection with high-$T_c$
superconductivity. Specifically, we elaborate on some work of Ivanov
and Sheka [14] concerning the dynamics of AFM vortices in a uniform
magnetic field $[15, 16]$.

The continuum dynamics of an antiferromagnet is described in terms of 
an order parameter $\nb n$ of unit length $(\nb n^2=1)$ which satisfies
the differential equation
$$\nb n\times\nb f=0;\qquad 
\nb f=\ddot{\nb n}-\Delta\nb n+2(\nb h\times\dot{\nb n})+
(h^2+\alpha^2)n_3\nb e.\eqno (34)$$
Here the constant $\alpha^2$ is the strength of a crystal easy-plane
anisotropy and $\nb h=(0, 0, h)=h\nb e$ is a uniform magnetic field
applied along the symmetry axis. In the absence of an applied field
$(h=0)$ eq. (34) reduces to the relativistic nonlinear $\sigma$ model
extended to include anisotropy. The effect of the external field is
twofold; it breaks Lorentz invariance, through the term 
$2(\nb h\times\dot{\nb n})$ in eq. (34), and also induces an effective
easy-plane anisotropy of strength $h^2$. Finally it is useful to 
derive eq. (34) from an action principle, i.e.,
$$\nb f=-{\delta{\cal A}\over \delta\nb n},\qquad
{\cal A}=\int Ldx_1dx_2dt\eqno (35)$$
where ${\cal A}$ is the action and $L$ is the Lagrangian density
$$L={1\over 2}[\dot{\nb n}^2-(\partial_{\mu}\nb n\cdot\partial_{\mu}\nb n)]
+\nb h\cdot (\nb n\times\dot{\nb n})-{1\over 2}
(h^2+\alpha^2)n^2_3.\eqno (36)$$
Hence we are armed with all the necessary information to carry out the
general program of sect. 2.

Lagrangian (36) may be parametrized in terms of spherical variables
$(n_1=\sin\Theta \cos\Phi$, $n_2=\sin\Theta \sin\Phi$, $n_3=\cos\Theta)$
to yield two pairs of canonically conjugate fields:
$$\matrix{\Psi_1=\Theta ,\hfill & \Pi_1=\dot{\Theta},\hfill\cr
\noalign{\bigskip}
\Psi_2=\Phi,\hfill & \Pi_2=(h+\dot{\Phi})\sin^2\Theta .\hfill\cr}
\eqno (37)$$
We may then insert these fields in the general expression for vorticity
given by eq. (3) to obtain
$$\gamma =\varepsilon_{\mu\nu}\partial_{\mu}(\dot{\nb n}\cdot\partial_{\nu}
\nb n)+h\omega ,\qquad
\omega\equiv{1\over 2}\varepsilon_{\mu\nu}\sin (2\Theta)\partial_{\mu}
(2\Theta)\partial_{\nu}\Phi .\eqno (38)$$
The first term is an uncomplicated total divergence and yields a vanishing
contribution to the total vorticity of eq. (9) which is then written as
$\Gamma =h\int\omega dx_1dx_2$. However the last integral may be different
from zero because $\omega$ resembles the Pontryagin density (23)
except for an overall factor $-{1\over 2}$ and the replacement
$\Theta\to 2\Theta$. The latter suggests considering the three-component
vector
$$N_1=2n_3n_1=\sin(2\Theta)\cos\Phi ,\quad
N_2=2n_3n_2=\sin (2\Theta)\sin\Phi ,\quad
N_3=2n^2_3-1=\cos (2\Theta),\eqno (39)$$
which is also a unit vector field $(\nb N^2=1)$. The density $\omega$
may be written as
$$\omega ={1\over 4}\varepsilon_{\mu\nu}(\partial_{\mu}\nb N\times
\partial_{\nu}\nb N)\cdot\nb N\eqno (40)$$
and resembles the standard Pontryagin density given in the third step of 
eq. (23). Furthermore the field $\nb N$ satisfies the simple boundary
condition $\nb N\to (0, 0, -1)$ at spatial infinity, thanks to the 
condition $n_3\to 0$ satisfied by all relevant configurations in the
presence of an easy-plane anisotropy. The net conclusion is that 
$\omega$ is actually the Pontryagin density for the field $\nb N$ and 
thus leads to an integer-valued total vorticity
$$\Gamma =h\int \omega dx_1dx_2=2\pi\kappa h;\qquad
\kappa =0, \pm 1, \pm 2, \ldots , \eqno (41)$$
which is nonvanishing when {\sl both} the applied field $h$ and the vortex
number $\kappa$ are different from zero. 

In the presence of either a crystal easy-plane anisotropy or an applied
field, or both, the relevant topological solitons are AFM vortices that
satisfy the boundary condition $n_3\to 0$ at spatial infinity. If one
insists on classifying these vortices by the standard winding number
of sect. 3, one would obtain $Q=-{1\over 2}\kappa\nu$ where
$\kappa =\pm 1$ is the vortex number and $\nu =\pm 1$ the polarity.
Hence $Q$ is half integer for AFM vortices which may thus be called
merons [17]. However the topological charge that is relevant for dynamics
is not $Q$ but the total vorticity $\Gamma =2\pi\kappa h$ which depends
on both the vortex number and the applied field but not on the polarity.
Therefore the general discussion of sect. 2 applied to the current example
suggests that AFM vortices should exhibit Hall or Magnus behavior only 
when an external field is present, and that the general dynamical picture
should be insensitive to the polarity.

The preceding general statement was thoroughly confirmed through
detailed numerical simulations $[15, 16]$. For $h=0$, two vortices with the
same vortex number $(\kappa_1=\pm 1=\kappa_2)$, initially at rest at a 
relative distance $d$, move off to infinity, just as two ordinary particles
would do when their interaction potential is repulsive. However, when a
nonvanishing uniform field $h$ is present, the two vortices actually
orbit around other, again in complete analogy with two interacting
electrons in a uniform magnetic field or two ordinary vortices in a
fluid. Similarly, when a vortex $(\kappa_1=1)$ and an antivortex
$(\kappa_2=-1)$ are initially at rest at a distance $d$, in the
absence of an external field, they converge toward each other and 
eventually annihilate. However, when a field is present, the 
vortex-antivortex pair undergoes Kelvin motion along two roughly parallel
lines that are perpendicular to the line connecting the vortex and the
antivortex. One should further note that above dynamical picture was
verified for either choice of relative polarity
$(\nu_1=1=\nu_2$ or $\nu_1=1=-\nu_2)$, also in agreement with the fact
that the total vorticity (41) is independent of polarity.

Therefore, to the extent that vortices are relevant for the physics of a 
2D antiferromagnet, the dynamical picture is changed significantly even 
by a very weak bias field. Perhaps the clearest manifestation of the 
effect of an applied field will emerge in the thermodynamics of an 
antiferromagnet. It is clear that much remains to be done in connection
with the anticipated Berezinskii-Kosterlitz-Thouless (BKT) phase 
transition that relies on the dynamics of a gas of vortices and antivortices.
Suffice it to say that the dynamics of vortex-antivortex pairs is radically
affected by the applied field. Hence the BKT theory may have to be 
reformulated in a way that clearly reflects the fundamental change of 
behavior when a field is turned on. Finally we mention that we have thus 
far confined our attention to the classical approximation. However it is
unlikely that the Hall behavior of classical AFM vortices described
here will be averted by quantum effects, especially because the overall
picture can be surmised directly from the conservation laws rather
than a detailed solution of the equations of motion. On the contrary,
one should expect that a full quantum treatment will lead to a 
richer picture, in analogy with the ``classical'' and ``quantum''
Hall effects of electrodynamics.

\bigskip
\bigskip
\noindent
{\twelvebf 5.$\,\,$ Superfluids}
\bigskip
We finally discuss briefly a class of problems pertaining to superfluids.
The simplest possibility is to consider the Hamiltonian dynamics 
defined from
$$i\dot{\psi}={\delta W\over\delta\psi^{\ast}},\qquad
i\dot{\psi}^{\ast}=-{\delta W\over\delta\psi},\eqno (42)$$
where $\psi =\psi (\nb x, t)$ is an order parameter, $\psi^{\ast}$
is its complex conjugate, and $W=W(\psi , \psi^{\ast})$ is some
energy functional. A typical choice for $W$ is the one that leads to the
Gross-Pitaevski model, often used as a simple model for superfluid
helium II [18].

A straightforward adaptation of the definition of vorticity given in 
eq. (20) yields
$$\bscal{\gamma} ={1\over i}(\bnabla\psi^{\ast}\times\bnabla\psi),
\eqno (43)$$
and the conserved linear and angular momenta are then obtained from the
general relations (21). Again, if partial integrations are performed
freely, these relations reduce to the canonical conservation laws. However
the latter are plagued by various ambiguities, as discussed by Jones
and Roberts [19] in their study of the dynamics of vortex rings as well
as of vortex-antivortex pairs. It is not difficult to see that the analysis
of the above reference can be repeated using the current definition
of conservation laws without encountering any ambiguities.

Actually the study of vortices also requires a 2D restriction of eq. 
(43) given by
$$\gamma ={1\over i}\varepsilon_{\mu\nu}\partial_{\mu}\psi^{\ast}
\partial_{\nu}\psi ,\eqno (44)$$
which is the direct analog of eq. (3), while the corresponding 
conservation laws are again given by eqs. (12) and (13). The associated
total vorticity reads
$$\Gamma =\int\gamma dx_1dx_2=2\pi\kappa ,\eqno (45)$$
where the vortex number $\kappa$ is an integer that can be extracted
also from the asymptotic behavior of the order parameter at spatial
infinity; $\psi\sim e^{i\kappa\phi}$. Therefore the general discussion
of sect.$\!$ 2 applies here without modification and, not surprisingly,
implies that the main dynamical features of superfluid vortices are
similar to those of ordinary vortices. This fact is confirmed by the
asymptotic analysis of Neu [20] which addresses the limit of widely
separated vortices. Furthermore the so-called Hall-Vinen drag induced by
the mutual friction between the superfluid and normal components can be
studied within an extended model introduced by Carlson [21], pretty
much along the lines of our discussion of skew deflection of ferromagnetic
bubbles in sect. 3.

Since the study of vortex dynamics in an ordinary fluid is usually
carried out in the idealized limit of an incompressible fluid [5], it is 
of interest to examine more closely our definition of vorticity for a 
superfluid which is inherently compressible. In particular, the superfluid
density vanishes at the location of a vortex and thus eliminates coordinate
singularities that would occur under the somewhat artificial assumption
of incompressibility. Now, if we write $\psi =\sqrt{\rho_s}\, e^{i\chi}$,
where $\rho_s=\psi^{\ast}\psi$ is the superfluid density, and further
identify the velocity field from $\nb u=\bnabla\chi$, the vorticity (43) may 
be written as $\bscal{\gamma}=\bnabla\times (\rho_s\nb u)$. This
expression differs from the standard definition of vorticity
$\bscal{\omega}=\bnabla\times\nb u$, except in the limit of an incompressible
fluid where $\rho_s=\hbox{const}$ and the two definitions differ only by an
overall constant, namely, $\bscal{\gamma}=\rho_s\bscal{\omega}$.
When the latter expression is inserted in the linear and angular momenta
of eq. (21), one recovers precisely the conservation laws for an ordinary
incompressible fluid given in eqs. (7.2.5) and (7.2.6) of ref. [5].

A related problem is that of the dynamics of Abrikosov vortices in a 
superconductor. Unfortunately this subject is controversial in that no
general agreement exists on a suitable phenomenological model for the 
description of the dynamics of the order parameter. Yet one should expect
that some general features of vortex dynamics do not depend on the details
of the model. In this spirit, a charged fluid was studied in refs. $[22, 23]$
that is described by a straightforward extension of the Gross-Pitaevski
model to include electromagnetism. The corresponding Abrikosov vortices
were then shown to exhibit dynamical properties analogous to those
encountered in all models mentioned in this review. In particular, 
unambiguous expressions for the linear and angular momenta were obtained
that are locally gauge invariant when expressed as moments of a suitable 
topological vorticity. One can further show that the same canonical 
structure persists in alternative models for a superconductor, such as a 
Chern-Simmons theory employed recently by Manton [24] to study the 
asymptotic dynamics of interacting Abrikosov vortices.

\bigskip
\bigskip
\noindent
{\twelvebf Acknowledgment}
\bigskip
I am grateful to my collaborators T.N. Tomaras, W.J. Zakrzewski, and
S. Komineas.

\bigskip
\bigskip
\noindent
{\twelvebf References}
\bigskip
\item{[1]\ }A.A. Thiele, Phys. Rev. Lett. 30, 230 (1973); 
J. Appl. Phys. 45, 377 (1974).
\smallskip
\item{[2]\ }A.P. Malozemoff and J.C. Slonczewski, Magnetic Domain Walls
in Bubble Materials (Academic Press, New York, 1979).
\smallskip
\item{[3]\ }N. Papanicolaou and T.N. Tomaras, Nucl. Phys. B 360, 425
(1991).
\smallskip
\item{[4]\ }N. Papanicolaou, in Singularities in Fluids, Plasmas and Optics,
eds. R.E. Caflisch and G.C. Papanicolaou (Kluwer, Amsterdam, 1993)
p. 151-158.
\smallskip
\item{[5]\ }G.K. Batchelor, An Introduction to Fluid Dynamics (Cambridge
University Press, Cambridge, 1967).
\smallskip
\item{[6]\ }A.A. Belavin and A.M. Polyakov, JETP Lett. 22, 245 (1975).
\smallskip
\item{[7]\ }N. Papanicolaou, Physica D 74, 107 (1994).
\smallskip
\item{[8]\ }N. Papanicolaou and W.J. Zakrzewski, Physica D 80, 225
(1995); Phys. Lett. A 210, 328 (1996).
\smallskip
\item{[9]\ }S. Komineas and N. Papanicolaou, Physica D 99, 81 (1996).
\smallskip
\item{[10]\ }B.I. Halperin and P.C. Hohenberg, Phys. Rev. 188, 898 (1969).
\smallskip
\item{[11]\ }D. Forster, Hydrodynamic Fluctuations, Broken Symmetry, and
Correlation Functions (Benjamin, Reading, 1975).
\smallskip
\item{[12]\ }S. Chakravarty, B.I. Halperin and D.R. Nelson, Phys. Rev.
B 39, 2344 (1989).
\smallskip
\item{[13]\ }V.G. Bar'yakhtar, M.V. Chetkin, B.A. Ivanov and S.N.
Gadetskii, Dynamics of Topological Magnetic Solitons - Experiment
and Theory (Springer Verlag, Berlin, 1994).
\smallskip
\item{[14]\ }B.A. Ivanov and D.D. Sheka, Phys. Rev. Lett. 72, 404 (1994).
\smallskip
\item{[15]\ }S. Komineas and N. Papanicolaou, Vortex Dynamics in 2D
Antiferromagnets, Crete preprint (1997).
\smallskip
\item{[16]\ }S. Komineas, PhD thesis, Crete (1997).
\smallskip
\item{[17]\ }D.J. Gross, Nucl. Phys. B 132, 439 (1978).
\smallskip
\item{[18]\ }R.J. Donnely, Quantized Vortices in Helium II (Cambridge
University Press, Cambridge, 1991).
\smallskip
\item{[19]\ }C.A. Jones and P.H. Roberts, J. Phys. A: Math. Gen. 15, 2599
(1982).
\smallskip
\item{[20]\ }J.C. Neu, Physica D 43, 385 (1990).
\smallskip
\item{[21]\ }N.N. Carlson, Physica D 98, 183 (1996).
\smallskip
\item{[22]\ }N. Papanicolaou and T.N. Tomaras, Phys. Lett. A 179,
33 (1993).
\smallskip
\item{[23]\ }G. Stratopoulos and T.N. Tomaras, Phys. Rev. B 54, 12493 (1996).
\smallskip
\item{[24]\ }N.S. Manton, Ann. Phys. (N.Y.) 256, 114 (1997).

\end